\documentstyle[preprint,revtex]{aps}
\tightenlines
\begin{document}
\draft
\hyphenation{Rijken}
\hyphenation{Nijmegen}
\hyphenation{NijmRdl}
\hyphenation{Stich-ting Fun-da-men-teel On-der-zoek Ma-te-rie}
\hyphenation{Ne-der-land-se Or-ga-ni-sa-tie We-ten-schap-pe-lijk}

\begin{title}
{\bf Comparison of potential models \\ with the
 \mbox{\boldmath $pp$} scattering data below 350 MeV}
\end{title}

\author{Vincent Stoks and J.J.\ de Swart}

\begin{instit}
Institute for Theoretical Physics, University of Nijmegen,
Nijmegen, The Netherlands\cite{email}
\end{instit}

\receipt{ }

\begin{abstract}
We calculate the $\chi^{2}$ of various $N\!N$ potential models with
respect to the $pp$ scattering data.
We find that only the potential models which were explicitly fitted
to the $pp$ data give a reasonable description of these data.
Most models give a pretty large $\chi^{2}$ on the very
low-energy $pp$ data, due to incorrect ${^{1}S}_{0}$ phase shifts.
\end{abstract}
\narrowtext

\nonum\section{{\bf I. INTRODUCTION}}
The $N\!N$ interaction has been the subject of investigation for more
than half a century, which is reflected in the numerous potential models
that have appeared in the literature. When discussing potential models
it has been convenient to divide the range of the
interaction into three regions~\cite{Tak51}:
a short-range, an intermediate-range, and a long-range region.
It soon became clear that in the long-range region ($r\agt2$ fm) the
$N\!N$ potential is given by one-pion exchange (OPE). The short-range
region ($r\alt1$ fm) is generally treated phenomenologically, where in
some models a form factor is introduced to make the potential regular
at the origin, whereas in other models a hard core is used.
For the description of the intermediate-range region ($1\alt r\alt2$ fm)
the first logical approach was to include the contributions from
two-pion exchange. Examples of the two-pion-exchange (TPE) potentials
of the early 1950s are the Taketani-Machida-Ohnuma~\cite{Tak52} and the
Brueckner-Watson~\cite{Bru53} models. However, these TPE models did not
give a satisfactory description of the $N\!N$ scattering data, mainly
due to a lack of sufficient spin-orbit force. The necessity of a
spin-orbit force was hinted at by Gammel, Christian, and
Thaler~\cite{Ga57a} when they failed to find a phenomenological
velocity-independent potential model consisting of central and tensor
parts which was able to fit all of the data available at that time.

The breakthrough came in 1957 with the simultaneous construction of the
purely phenomenological Gammel-Thaler potential~\cite{Ga57b} and the
semi-phenomenological Signell-Marshak potential~\cite{Sig58}, both
models introducing phenomenological spin-orbit potentials.
The Gammel-Thaler model gave a good fit to the scattering data
up to 310 MeV. The Signell-Marshak model, consisting of the TPE
Gartenhaus potential~\cite{Gar55} together with a phenomenological
spin-orbit force, was successful up to 150 MeV.
These potential models were soon improved upon, where we mention the
hard-core Hamada-Johnston~\cite{Ham62} and Yale~\cite{Las62} models,
and the various hard-core and soft-core models constructed by
Reid~\cite{Rei68}.

With the discovery of the heavy mesons in the early 1960s a different
approach was launched. In this approach the $N\!N$ interaction is
written as a sum over one-boson-exchange (OBE) potentials. The OBE
models are very successful in that the vector-meson and scalar-meson
exchanges supply the required spin-orbit forces.
However, the short-range part still has to be described
phenomenologically. Examples of some of the early OBE models are
given in Refs.~\cite{Hos61,Won64,Bry64}, whereas some of the more
recent models will be discussed in Sec.\ III.

Most of the early models gave a for that time reasonable description
of the $N\!N$ scattering data. However, over the years the concept over
what is accepted as reasonable has changed. At present, a model which
describes the $pp$ scattering data will only be called reasonably good
if it has $\chi^{2}/N_{\rm data}\alt2$, where $N_{\rm data}$ denotes
the number of $pp$ scattering data. In Sec.\ III we show that only few
of the models that have appeared in the literature still live up to
this criterion of having $\chi^{2}/N_{\rm data}\alt2$.
On the other hand, with the fast computers and the accurate partial-wave
analyses presently available, it has now become rather easy to construct
phenomenological potential models which have the excellent
quality of $\chi^{2}/N_{\rm data}\approx1$. Such a potential model for
the $N\!N$ interaction with $\chi^{2}/N_{\rm data}\approx1$ is the best
one can hope to achieve. The quality of the fit of such a potential
model can compete with the fit of a multienergy partial-wave analysis,
and in a sense provides another form of partial-wave analysis.
Several of such high-quality potential models will be presented in a
forthcoming paper~\cite{St92b}.

In this paper we will use the Nijmegen representation of the
$\chi^{2}$ hypersurface of the $pp$ scattering data to calculate the
$\chi^{2}$ with respect to the $pp$ scattering data for a number of
different $N\!N$ potential models. The Nijmegen representation is
obtained from the single-energy analyses of the Nijmegen $pp$
partial-wave analysis~\cite{Ber90}. It provides an adequate
representation of the scattering data.

We want to stress that such a comparison between potential models can
only be done fairly when for {\it all} potential models the $\chi^{2}$
with respect to the data is calculated correctly and in the same way.
This in order to avoid any ambiguities regarding whether or not any
specific electromagnetic corrections are accounted for. For that purpose
we calculate the phase shifts for each potential model by solving
the radial Schr\"odinger equation in the presence of the Coulomb
interaction. For the ${^{1}S}_{0}$ phase shift the vacuum polarization
potential and the two-photon-exchange modification to the Coulomb
potential are also included (for details see, e.g., Ref.~\cite{Ber90}).
We make one exception, however: For the momentum-space Bonn
potential~\cite{Hai89} we use the phase shifts as provided by one
of the authors. The reason is that a proper treatment of the
electromagnetic interaction is very important if the $pp$ scattering
data are to be described properly (see, e.g., Ref.~\cite{Sto90}), and
we do not have the software programs for handling electromagnetic
effects in momentum-space calculations. For $T_{\rm lab}<30$ MeV,
the ${^{1}S}_{0}$ phase shifts of the Bonn potential were adjusted
by us such as to account for vacuum polarization and modified Coulomb
effects. The proper way of how to make these adjustments is described
in Ref.~\cite{Ber88}.

Another important aspect for a fair comparison between various potential
models is that they should all be compared with the same database. We
use the database of the Nijmegen partial-wave analysis~\cite{Ber90}.
This database is as complete as possible and has been scrutinized very
carefully where all bad data or groups of bad data have been removed.

In Sec.\ II we explain how the Nijmegen representation of the
$\chi^{2}$ hypersurface of the $pp$ scattering data can be used to
calculate the $\chi^{2}$ with respect to these data for any particular
potential model. The advantage of using the Nijmegen representation is
that the phase shifts of a model that is to be investigated need only
be calculated at a small number of energies, namely at the 10
energies of the single-energy analyses of the Nijmegen partial-wave
analysis~\cite{Ber90}. This small number (10) should be compared with
the much larger number of energies at which experimental data have been
measured (about 200). Therefore, using the Nijmegen representation
of the $pp$ data rather than the data themselves saves a lot of
computer time, while the results are almost always sufficient for
their purpose.

In Sec.\ III we present our results for some of the more well-known
potential models that have appeared in the literature, most of which
are still commonly used in other calculations involving the $N\!N$
interaction. Examples of such calculations are $pp$ bremsstrahlung,
three-nucleon elastic scattering, few-nucleon bound-state calculations,
and nuclear matter calculations.
The list of potential models that we discuss is not complete. For
example, momentum-space potentials will not be considered here, due
to reasons already discussed above. The only exception is the
momentum-space Bonn potential, since it is widely used and a specific
$pp$ version has been published~\cite{Hai89}.
Four of the more recent potential models which give a good description
of the $pp$ data are then studied in more detail.

\nonum\section{{\bf II. REPRESENTATION OF $\chi^{2}$ HYPERSURFACE}}
The details of the Nijmegen way of analyzing the $pp$ scattering data
are extensively discussed elsewhere~\cite{Ber88,Ber90}, and will not
be repeated here. Here we want to stress that the Nijmegen
partial-wave analysis is not particularly important for the present
calculations, because we compare various potential models with the
experimental data. What is important is that we can use the same
computer programs as used in the Nijmegen partial-wave analyses to
compute properly the $pp$ scattering amplitudes. In doing the comparison
we make use of a representation of the $\chi^{2}$ hypersurface of the
$pp$ scattering data, which was produced by the Nijmegen single-energy
partial-wave analyses of the $pp$ scattering data. This $\chi^{2}$
hypersurface is somewhat dependent on the Nijmegen multienergy (m.e.)
analysis, but the crucial point is that it provides an excellent
representation of the $pp$ data, as will be demonstrated below.

The Nijmegen representation of the $\chi^{2}$ hypersurface is
obtained as follows. In the single-energy (s.e.) analyses the 1787 $pp$
scattering data below $T_{\rm lab}=350$ MeV are clustered at 10
energies from 382.54 keV (the interference minimum) up to 320 MeV.
The total $\chi_{\rm se}^{2}$ of all 10 s.e.\ analyses amounts to
$\chi_{\rm se}^{2}=1676.3$. These s.e.\ analyses provide us with 10
error matrices $E_{n}$. The error matrix is the inverse of half
the second-derivative matrix of the $\chi^{2}$ hypersurface with respect
to the phase shifts within a particular energy bin. The Nijmegen
representation of the $\chi^{2}$ hypersurface of the $pp$ scattering
data consists of the number $\chi^{2}_{\rm se}=1676.3$ and the 10
error matrices $E_{n}$ at the 10 different energies. It provides a
good representation of the scattering data within each energy bin.
However, this representation is not exact. First of all, the higher
partial-wave phase shifts are fixed at their m.e.\ values.
Furthermore, the data have been clustered at some central energy
within an energy bin using the results of the m.e.\ fit, and next to
that we have used the approximation that the $\chi^{2}$ hypersurface
is quadratic in the neighborhood of the minimum.

The representation of the data can be used as follows.
The phase shifts of some model are calculated at the 10 central
energies of the s.e.\ analyses. Denoting by ${\bf d}_{n}$ the
deviation of the phase-shift predictions of the model from the s.e.\
phase shifts in the $n$th energy bin, the $\chi^{2}$ of the model can
be written as a sum of the s.e.\ contributions $\chi_{{\rm se},n}^{2}$
and the contributions from the inverse error matrices
$\chi_{{\rm rep},n}^{2}$, i.e.,
\begin{eqnarray}
    \chi^{2}({\rm model}) &=& \sum_{n}\left(\chi^{2}_{{\rm se},n}+
       \chi^{2}_{{\rm rep},n}\right)                         \nonumber\\
      &=& \chi^{2}_{\rm se} +
     \sum_{n} {\bf d}^{T}_{n} E_{n}^{-1} {\bf d}_{n}  \ . \label{chimod}
\end{eqnarray}
In using Eq.~(\ref{chimod}), we account for the correlations between
the different phase shifts, because this information is stored
in the error matrices.

In order to investigate the quality of the Nijmegen representation of
the $\chi^{2}$ hypersurface, we tested it in several ways. First, we
used our m.e.\ phase shifts as model phase shifts and
calculated the corresponding $\chi^{2}$ contribution. These
$\chi_{{\rm rep},n}^{2}$ contributions of the m.e.\ phase shifts
are listed in the last column of Table~\ref{pwa}. They should be added
to the $\chi_{{\rm se},n}^{2}$ of the s.e.\ analyses given in the next
to last column of Table~\ref{pwa} to give the total $\chi^{2}$ within
each energy bin. For all 10 energy bins we find that
the agreement of $\chi_{{\rm se},n}^{2}+\chi_{{\rm rep},n}^{2}$ with
the corresponding $\chi_{{\rm me},n}^{2}$ of the m.e.\ analysis is
very good.
The difference between the total $\chi^{2}({\rm model})=1786.8$ given
by Eq.~(\ref{chimod}) and the $\chi_{\rm me}^{2}=1786.4$ reached in our
m.e.\ analysis is only 0.4. This means that the $\chi^{2}$ as calculated
directly on the data and the $\chi^{2}$ calculated via
Eq.~(\ref{chimod}) only differ by 0.02\%.
It shows that the approximation, that the $\chi^{2}$ hypersurface of
the s.e.\ analyses is quadratic up to the minimum $\chi^{2}_{\rm me}$
of the m.e.\ analysis, is actually very good.
For completeness, we have also listed in Table~\ref{pwa} the number of
scattering data $N_{\rm data}$ within each energy bin, which is the
number of scattering observables plus the number of normalizations with
an experimental error. The information presented in Table~\ref{pwa}
forms the basis for our test of the quality of various potential
models to be discussed in the sections below.

As a second test for the quality of the Nijmegen representation of the
$\chi^{2}$ hypersurface, we used the Nijmegen soft-core potential
(Nijm78)~\cite{Nag78} to compare the $\chi^{2}({\rm model})$ obtained
using Eq.~(\ref{chimod}) with the $\chi^{2}({\rm data})$ obtained
from a direct comparison with the data. We are now farther away from
the minimum $\chi^{2}$, so we expect that the $\chi^{2}$ hypersurface
will no longer be quadratic. As a consequence, the result for
$\chi^{2}({\rm model})$ using Eq.~(\ref{chimod}) will be less accurate.
For the Nijm78 model we find $\chi^{2}({\rm data})=3387.5$ and
$\chi^{2}({\rm model})=3462.8$. The difference of 75.3 is now about
2\%, which is sufficiently small. When we are farther away from
the minimum $\chi^{2}$, this difference will be even larger, but
Eq.~(\ref{chimod}) is still correct within the order of magnitude.
This allows us to use Eq.~(\ref{chimod}) to make statements regarding
the quality of some potential model.

The difference between $\chi^{2}({\rm model})$ and $\chi^{2}({\rm data})$
can be understood as follows.
For the calculation of the $\chi^{2}$ via Eq.~(\ref{chimod}) only the
lower partial waves of the potential model up to $J=4$ are used.
All higher partial waves are taken from the m.e.\ analysis. Also all
normalization constants are fixed at the values as obtained in the
s.e.\ analyses. On the other hand, in the direct comparison with the
data for the calculation of $\chi^{2}({\rm data})$ all partial waves
of the potential model up to $J=8$ are used. Furthermore, all
normalization constants are automatically adjusted such as to give
the best agreement with the data. This makes that
$\chi^{2}({\rm data})$ is smaller than $\chi^{2}({\rm model})$.

\nonum\section{{\bf III. COMPARISON \\ OF NN POTENTIAL MODELS}}
The Nijmegen representation of the $\chi^{2}$ hypersurface of the $pp$
scattering data can be used to test the quality of an $N\!N$ potential
model. For that purpose we calculate the $\chi^{2}({\rm model})$ using
Eq.~(\ref{chimod}). The results for a number of models are shown in
Table~\ref{chipot}, where we present the $\chi^{2}_{{\rm rep},n}$
contributions. To obtain the total $\chi^{2}$ within a particular
energy bin, one should add the $\chi^{2}_{{\rm se},n}$ as given in
Table~\ref{pwa}.
As discussed in the previous section, the larger entries
($\agt$ 1000) in Table~\ref{chipot} are inaccurate, but they still
correctly represent a large $\chi^{2}_{{\rm rep},n}$.
Before discussing each of the models in more detail we first note
some general features.

For some models the 50.0 MeV energy bin gives a relatively large
contribution to $\chi^{2}$. This is partially due to a recently
published very accurate analyzing power experiment at 50.04
MeV~\cite{Smy89}. The accuracy of this 50.04 MeV experiment makes
that especially the triplet $P$ phase shifts at 50 MeV are now very
accurately known.
Because the triplet $P$ phase shifts of the various potential models
are not always in too good an agreement with these new accurate values,
they will produce a large contribution to $\chi^{2}$.

Many models give a relatively poor or even very bad description of the
low-energy data. The reason is that the ${^{1}S}_{0}$ phase shifts at
382.54 keV and 1.0 MeV are very accurately known. So when the
${^{1}S}_{0}$ phase shift of a potential model is a little bit off,
the $\chi^{2}$ contribution will already be enormous.
However, there still remains the fact that most of these models claim
to fit the scattering length and the effective range pretty well, so
one would expect that the $\chi^{2}$ contribution of the low-energy data
should not be too large (or as large as it is for some of the models).
In order to investigate whether the high $\chi^{2}$ value for some
of the models is only due to an erroneous ${^{1}S}_{0}$ phase shift
at these lowest energies, we also compared the various models in a
slightly different energy range.
In Table~\ref{chipot} we therefore also give the $\chi^{2}$ contribution
on the 2--350 MeV energy range, which contains 1590 $pp$ scattering
data. If now the quality of a potential model is bad, it is not totally
due to a slightly incorrect ${^{1}S}_{0}$ phase shift at low energies.

In the following we chronologically list some of the better-known
$N\!N$ potentials that have appeared in the literature, most of
which are still commonly used in other calculations involving the
$N\!N$ interaction. The older phenomenological and TPE models of the
early 1950s are not included.

-- HJ62: {\it Hamada-Johnston potential}~\cite{Ham62} \\
The energy-independent Hamada-Johnston potential is a hard-core
potential. It includes the OPE potential and a phenomenological part
consisting of central, tensor, spin-orbit, and quadratic spin-orbit
terms. At the time of its presentation it provided a good
representation of the $pp$ and $np$ scattering data below 315 MeV.
The 28 model parameters were fitted to the Yale phase
shifts~\cite{Bre62}. In 1970, Humberston and Wallace~\cite{Hum70}
introduced an additional parameter to improve the deuteron properties
of the model.
{}From Table~\ref{chipot} we see that the data in the 50 MeV bin and
the very low-energy data give a large contribution to $\chi^{2}$.
This is not surprising in view of the much higher accuracy with which
these phase shifts are known nowadays.
For the description of the remaining 1347 data this old HJ62 model
is still surprisingly good, but it is only sparsely used anymore.

-- Reid68: {\it Reid soft-core potential}~\cite{Rei68} \\
In the paper by Reid, a number of different hard-core and soft-core
potentials are presented. In these models each partial wave with
total angular momentum $J\leq2$ is parametrized phenomenologically
in terms of Yukawa functions of multiples of the pion mass. The OPE
part itself is explicitly included. In some partial waves an explicit
distinction between central, tensor, and spin-orbit parts is used.
A shortcoming of the soft-core versions is that the potentials are
not regular in the origin, but still have an $r^{-1}$ singularity.
The parameters were fitted to the $pp$ and $np$ phase shifts of
the Yale~\cite{Bre62} and early Livermore~\cite{Arn66} analyses.
In 1981, Day~\cite{Day81} extended the potential for partial waves
with $J>2$.
Also for this model the description of the very low-energy data is a
bit off. The remaining data are described pretty well.

-- TRS75: {\it Super-soft-core potential}~\cite{Tou75} \\
This $pp+np$ potential contains the $\pi$-, $\rho$-, and
$\omega$-exchange contributions where the coupling constants are taken
from other sources. The other important intermediate-range
contributions to the $N\!N$ force are parametrized phenomenologically
through OBE potential functions with 32 free ranges and amplitudes.
The potential contributions are regularized at the origin by step-like
functions which also serve to construct the short-range
phenomenological cores, whence the name super-soft-core potential.
The model is an improved version of an earlier super-soft-core model
by the same group~\cite{Tou73}.
The model is very good for the 0.5--35 MeV energy region, whereas for
higher energies the description rapidly becomes worse. Also the very
low-energy data of the 0.38254 MeV bin are not described very well,
even though the pion and nucleon masses used in the $pp$ and $np$
potentials were especially adjusted so as to account for the
difference between the $pp$ and $np$ ${^{1}S}_{0}$ phase shifts.

-- OBEG75: {\it Funabashi potentials}~\cite{Obi75} \\
These potentials are constructed from the $\pi, \eta, \rho$, and
$\omega$ OBE potentials. Also included are the contributions of two
scalar mesons $\delta$ and $\sigma$, the masses of which were fitted
to the scattering data. The potential contains the standard OBE part
and a retardation part. The off-energy-shell effects coming from the
retardation, albeit of little importance to the two-nucleon system,
are expected to play an important role in many-nucleon systems.
The potentials were evaluated in coordinate space for the sake of
future investigations regarding the influence of off-energy-shell
effects in finite nuclei.
The various treatments of the inner region in these potentials are
a hard core, a Gaussian soft core, and a velocity-dependent core.
In each case an attractive spin-orbit core is included to improve the
triplet $P$ phase shifts. Furthermore, all potentials are regularized
by means of a step-like cutoff function. The results presented here
refer to the Gaussian soft-core potential, denoted by OBEG.
{}From Table~\ref{chipot} we see that the overall behavior of this
model is rather bad.

-- Nijm78: {\it Nijmegen potential}~\cite{Nag78} \\
The mesons which give rise to the meson-exchange forces of the Nijmegen
potential are the non-strange mesons of the pseudoscalar, vector, and
scalar nonets. They can be identified with the dominant parts of the 9
lowest-lying meson trajectories in the complex $J$ plane.
The model also includes the dominant $J=0$ parts of the Pomeron,
$f, f', A_{2}$ trajectories, which essentially lead to repulsive central
Gaussian potentials.
The inner region is adjusted with an exponential form factor. The 13
model parameters were fitted to the phase-shift error matrices of
the 1969 Livermore analyses~\cite{Mac69}. These model parameters
can be checked with meson-nucleon coupling constants and cutoffs
obtained from other sources. An important feature of this model is
that there exist exactly equivalent versions of this potential for use
in coordinate space or momentum space. Using the same set of parameters,
both the coordinate-space and momentum-space versions produce
the same phase shifts at all energies (see also Ref.~\cite{Rij91}).
The overall description of the $pp$ data is good; only the $\chi^{2}$
contribution to the 50 MeV bin is a little bit high.

-- Paris80: {\it Parametrized Paris potential}~\cite{Lac80} \\
The original Paris potential~\cite{Cot73} was obtained by calculating
the TPE contributions to the $N\!N$ forces from the pion-nucleon phase
shifts and from the pion-pion interaction using dispersion relations.
The $\pi$- and $\omega$-exchanges were then also explicitly included.
A balanced fitting to the phase-shift error matrices of the 1969
Livermore analysis~\cite{Mac69} and to the $pp$ and $np$ scattering
data themselves required a total of 12 parameters. In 1980 a
parametrized version~\cite{Lac80} consisting of a set of Yukawa
functions provided a phenomenological representation of the Paris
potential.
Except for the very low-energy region, this model gives a good
description of the $pp$ scattering data, where also the description
of the 50 MeV bin is not too bad.

-- Urb81: {\it Urbana potential}~\cite{Lag81} \\
The Urbana potential is a purely phenomenological $v_{14}$ potential
where 14 represents the number of different potential types (central,
spin-spin, tensor, spin-orbit, quadratic spin-orbit, centrifugal,
centrifugal spin-spin, and an overall isospin dependence), rather
than the number of phenomenological parameters. Next to OPE
and a 14-parameter representation of TPE, the short-range part is
represented by two Woods-Saxon potentials using a total of 20
parameters. All potential types are regularized by means of a cutoff
function. The parameters were fitted to the $np$ phase shifts of the
1977 energy-dependent phase shift analysis by Arndt
{\it et al.}~\cite{Arn77}.
The 50 MeV bin and the 150 MeV bin give relatively large
contributions to $\chi^{2}$.
Also here the description of the very low-energy data is a bit off.

-- Arg84: {\it Argonne potential}~\cite{Wir84} \\
The Argonne potential is similar to the Urbana potential. It was
fitted to a 1981 phase shift analysis of Arndt and Roper (an update
of the analysis of Ref.~\cite{Arn77}) for the $np$ scattering data in
the 25--400 MeV energy range. Next to OPE and a 14-parameter
representation of TPE, the short-range part of the Argonne potential
is represented by a Woods-Saxon potential using 16 parameters. The
main reason for constructing this new $v_{14}$ model was to have a
phase-equivalent standard of comparison for the $v_{28}$ model, which
includes operators which represent all possible processes with
$N\!\Delta\pi$ or $\Delta\Delta\pi$ vertices.
The description of the very low-energy data is bad, which is
not surprising in view of the fact that the model was fitted
to the $np$ data with $T_{\rm lab}>25$ MeV. Also the 50 MeV
and 150 MeV bin are not described too well. Still, in the 25--350 MeV
region the Argonne model provides an improvement over the
Urbana model.

-- Bonn87: {\it Coordinate-space Bonn potential}~\cite{Mac87} \\
The full Bonn potential is an $N\!N$ momentum-space potential. Next to
$\pi$-, $\omega$-, and $\delta$-exchanges, the model also contains an
explicit determination of the TPE contribution, including
$\rho$-exchange and virtual isobar excitation. The combined
$\pi\rho$-exchange diagrams are included as well.
The coordinate-space version is obtained from a simple parametrization
of the full model by 6 OBE terms (three pairs of pseudoscalar, vector,
and scalar mesons, respectively). The potentials are regularized at the
origin by means of dipole form factor functions.
In this paper we use the coordinate-space OBE version in order to be
able to include the electromagnetic interaction.
We find that the description of the very low-energy data is very bad,
while for the higher energies the description is not too good either.
It demonstrates that the coordinate-space Bonn potential is only a poor
substitute for the full Bonn model, which is claimed to give a
reasonably good description of the scattering data~\cite{Mac87}. There
have also appeared a number of other adjusted OBE coordinate-space
versions~\cite{Mac89}, Bonn A and Bonn B, which also give a very poor
description of the $pp$ data ($\chi^{2}/N_{\rm data}>8$ in the 2--350
MeV energy range).

-- Bonn89: {\it Updated Bonn potential}~\cite{Hai89} \\
This potential model is an adaptation of the full momentum-space Bonn
potential~\cite{Mac87} to the $pp$ scattering data. This was done by
including the Coulomb interaction in the momentum-space calculations
and making small adjustments to guarantee a reasonable confrontation
with the $pp$ data. The scalar-meson coupling constants were changed
in such a way as to explicitly fit the $pp$ ${^{1}S}_{0}$ phase shift
below 2 MeV (as suggested in Ref.~\cite{Ber88}), while keeping the
deuteron properties in the $np$ ${^{3}S}_{1}-{^{3}D}_{1}$ channel at
the values of the original full Bonn model.
Indeed, now the overall description of the $pp$ data is good.

-- NijmRdl: {\it Reidlike Nijmegen potential}~\cite{St92b} \\
This model is an example of a new class of high-quality potential
models, which are almost as good as the Nijmegen m.e.\ $pp$ partial-wave
analysis~\cite{Ber90}. The model is a Reidlike version of the Nijm78
model~\cite{Nag78} in the sense that we define a potential form for
each partial wave separately. For each partial wave we only need to
adjust a few parameters of the original Nijm78 model in order to
arrive at a semi-phenomenological potential model which gives an
excellent description of the scattering data. In that sense this
Reidlike potential model is another form of m.e.\ partial-wave analysis.
Preliminary versions of this Reidlike Nijmegen model were presented
at the Shanghai and Elba conferences~\cite{Swa90,Swa91}.
Comparison of the last column of Table~\ref{pwa} with the last column
of Table~\ref{chipot} clearly demonstrates the excellent quality of
this potential model.

Summarizing, only the Nijm78 and Bonn89 potentials give a rather good
description of the $pp$ scattering data over the entire 0--350 MeV
energy range. When we do not include the very low-energy (0--2 MeV)
data, also the Reid68 and Paris80 models are reasonably good, as can
be read off from the last line of Table~\ref{chipot}.
These four models are then roughly of the same quality, i.e.,
$\chi^{2}/N_{\rm data}\approx2$.
However, these ``good'' models are still not as good as the NijmRdl
model which has $\chi^{2}/N_{\rm data}\approx1$, which is very close
to the $pp$ partial-wave analysis.
We therefore believe that one has to be very careful in drawing
conclusions regarding the importance or unimportance of, e.g.,
three-nucleon forces in many-body calculations, when these conclusions
are only based on calculations where the $N\!N$ interaction is
represented by an $N\!N$ potential model which cannot even adequately
describe the two-nucleon scattering data.

In the remaining part of this section we will focus on the four
recent potential models which give a good (Nijm78, Paris80, Bonn89)
or excellent (NijmRdl) description of the $pp$ scattering data.
It is very instructive to see how the different partial waves contribute
to the total $\chi^{2}$. For that purpose we start with the m.e.\ phase
shifts and substitute the ${^{1}S}_{0}$ phase shifts of the
different potential models. We then calculate the difference
${\scriptstyle\Delta}\chi^{2}$ between this new $\chi^{2}_{\rm rep}$
and the $\chi^{2}$ of the m.e.\ analysis. This is repeated for the other
lower partial-wave phase shifts up to $J=3$ as well. In this way we
can judge the quality of the various partial waves of these models.
The six separate contributions can be summed and compared with
the $\chi^{2}$ as obtained when we take all these potential phase
shifts simultaneously as given in Table~\ref{chipot}, which gives
some measure for the importance of the correlation between the
different partial waves.
The results are presented in Table~\ref{potwave}.
The agreement between the sum of the ${\scriptstyle\Delta}\chi^{2}$
contributions substituting the potential model phase shifts one at a
time, and the ${\scriptstyle\Delta}\chi^{2}$ contribution using all
potential model phase shifts simultaneously is not very good, the
result for the Bonn89 potential being the worst. This means that the
correlation between the various phase shifts in the Bonn89 potential
is very important.

The ${^{3}P}_{1}$ phase shift of the Nijm78 potential is found to
be very close to the m.e.\ value. For the other phase shifts, the
${\scriptstyle\Delta}\chi^{2}$ contributions are about the same
for each of the separate contributions. The disagreement between the
Nijm78 potential and the m.e.\ analysis is largest for the
${^{1}D}_{2}$ phase shift.

For the Paris80 potential the ${^{1}D}_{2}$ and coupled
${^{3}P}_{2}-{^{3}F}_{2}$ phase shifts are not too good,
whereas the other phase shifts are in reasonable agreement with the
m.e.\ analysis.

Similarly, for the Bonn89 potential the ${^{1}S}_{0}$, the
${^{3}P}_{0}$, and the coupled ${^{3}P}_{2}-{^{3}F}_{2}$ phase shifts
are not very good. This is partially due to the fact that the isovector
tensor force of the Bonn89 potential is too strong and its spin-orbit
force is too weak, which can be concluded from comparing the tensor
and spin-orbit combinations of the $^{3}P$ phase shifts with the
corresponding combinations as obtained in the $pp$ partial-wave
analysis.

The ${\scriptstyle\Delta}\chi^{2}$ differences of the NijmRdl model
are much smaller. For the ${^{1}D}_{2}$ phase shift the difference
is even negative, which means that for this partial wave the NijmRdl
model is better than the m.e.\ analysis. Also the ${^{3}F}_{3}$
partial wave is slightly better than in the m.e.\ analysis.

\nonum\section{{\bf IV. CONCLUDING REMARKS}}
We have tested the quality of a number of $N\!N$ potentials with
respect to the $pp$ scattering data in the 0--350 MeV energy range.
Of the older models only the Reid68, Nijm78, and Paris80 models give
satisfactory results when confronted with the $pp$ data.
The new Bonn89 model, an adjustment of the full Bonn potential to
explicitly fit the $pp$ data, is of a similar quality as the
Nijm78 and Paris80 potentials in the 2--350 MeV energy range.

If we also include the very low-energy data (0--2 MeV), only the
Nijm78 and Bonn89 potentials still give a reasonable description of
the data. The other models all give a large to very large contribution
to $\chi^{2}$ in this low-energy region. The reason is that the $pp$
${^{1}S}_{0}$ phase shift at $T_{\rm lab}=382.54$ keV is very accurately
known. So a small deviation for the ${^{1}S}_{0}$ prediction from one
of these potential models will give rise to an enormous contribution to
$\chi^{2}$. However, this contribution should not be too large, since
most potential models claim to give a good description of the
scattering length and effective range parameters.
Furthermore, the fact that some of the models give a rather poor
description of the $pp$ data is not only due to an incorrect
${^{1}S}_{0}$ phase shift. As an example we consider the Arg84
potential. In the 2--350 MeV energy range the Arg84 model gives
$\chi^{2}/N_{\rm data}=7.1$. When we replace the Arg84 ${^{1}S}_{0}$
phase shifts by our m.e.\ values (which roughly corresponds to having
a model with ``perfect'' ${^{1}S}_{0}$ phase shifts), the quality of
the model improves considerably. However, the resulting
$\chi^{2}/N_{\rm data}\approx4$ is still rather large.
This demonstrates that the other phase shifts are not too good either.

An important conclusion which can be drawn from the potential
comparison with the $pp$ scattering data discussed in this paper is
that only the potential models which were explicitly fitted to the $pp$
data (Nijm78, Paris80, Bonn89) give a reasonable description of these
data. Here we have to keep in mind that the Nijm78 and Paris80 models
were fitted to the 1969 Livermore database~\cite{Mac69}. Our present
database contains a large number of new and more accurate data, which
are still described rather well by these two models.
The Bonn89 potential was fitted to a much more recent database,
not too different from our present database.
Apparently, a good fit to the $np$ data does not automatically
guarantee a good fit to the $pp$ data. One of the reasons is that the
$np$ data are less accurate than the $pp$ data, so the constraints on
the $np$ phase shifts are not so large. Also, the difference between
the $pp$ and $np$ ${^{1}S}_{0}$ phase shifts should be included
explicitly.

We therefore conclude that if a potential model is claimed to give a
good description of the $pp$ scattering data, this claim should be
based on an explicit confrontation of the model with these $pp$ data,
either directly or using Eq.~(\ref{chimod}).

Part of this work was included in the research program of the Stichting
voor Fundamenteel Onderzoek der Materie (FOM) with financial support
from the Nederlandse Organisatie voor Wetenschappelijk Onderzoek (NWO).

\newpage
\narrowtext
\begin{table}
\caption{The $\chi^{2}$ results of the $pp$ partial-wave analyses
         for the 10 single-energy bins.}
\begin{tabular}{rr@{--}lrrrr}
  energy & \multicolumn{2}{c}{bin} & $N_{\rm data}$
         & $\chi^{2}_{{\rm me},n}$ & $\chi^{2}_{{\rm se},n}$
         & $\chi^{2}_{{\rm rep},n}$                      \\
  \tableline
  0.38254 &   0.0&0.5  &  134  &  134.5  &  129.2  &   5.3  \\
      1.0 &   0.5&2    &   63  &   39.7  &   37.4  &   2.3  \\
      5.0 &     2&8    &   48  &   44.6  &   30.9  &  13.9  \\
     10.0 &     8&17   &  108  &  102.9  &   87.8  &  14.9  \\
     25.0 &    17&35   &   59  &   63.1  &   62.0  &   1.1  \\
     50.0 &    35&75   &  243  &  212.9  &  206.4  &   6.6  \\
    100.0 &    75&125  &  167  &  170.8  &  150.8  &  19.7  \\
    150.0 &   125&183  &  343  &  377.9  &  356.7  &  21.5  \\
    215.0 &   183&290  &  239  &  286.1  &  265.8  &  20.7  \\
    320.0 &   290&350  &  383  &  353.7  &  349.3  &   4.5  \\
  \tableline
          &     0&350  & 1787  & 1786.4  & 1676.3  & 110.5  \\
          &     2&350  & 1590  & 1612.2  & 1509.8  & 102.8  \\
\end{tabular}
\label{pwa}
\end{table}

\widetext
\begin{table}
\caption{The $\chi^{2}_{{\rm rep},n}$ results at the 10 single-energy
         bins for various $N\!N$ potential models. The short-hand
         notation for each model is defined in Sec.\ III.
         In order to arrive at the total $\chi^{2}$ one has to add
         the $\chi_{{\rm se},n}^{2}$ contributions of the analyses listed
         in Table~\protect\ref{pwa}. The $\chi^{2}/N_{\rm data}$ in the
         bottom line refers to the data in the 2--350 MeV energy range.}
\begin{tabular}{r@{--}l|rrrrrrrrrrr}
  \multicolumn{2}{c|}{bin}  & HJ62  & Reid68  & TRS75  & OBEG75  & Nijm78
     & Paris80  & Urb81  & Arg84  & Bonn87  & Bonn89  & NijmRdl \\
  \tableline
 0.0&0.5 &  6620 &  880 &  480 &  25500 &   62 & 3660
         &   980 & 845000 & 665000 &  71 &   5.3 \\
 0.5&2   &  1960 &  132 &  100 & 610000 &    8 &  773
         &    20 & 230000 & 195000 &   7 &   2.3 \\
   2&8   &    29 &   63 &   20 &   5370 &   51 &   18
         &   115 &   1960 &   2400 &   8 &  15.2 \\
   8&17  &   103 &  206 &   55 &   3980 &   76 &   33
         &   275 &   1470 &   2540 &  46 &  13.6 \\
  17&35  &   201 &    9 &   23 &    960 &   67 &   13
         &   575 &    675 &   1950 &  20 &   2.1 \\
  35&75  &  6370 &  300 &  980 &   5330 &  555 &  333
         &  1920 &   1365 &   6090 & 346 &   8.0 \\
  75&125 &   110 &  128 &  332 &    320 &  131 &   41
         &   470 &    265 &    840 &  57 &  26.1 \\
 125&183 &   305 &  242 &  630 &   6540 &  222 &  415
         &  3280 &   3060 &   1870 & 284 &  17.9 \\
 183&290 &   227 &  110 &  980 &   2750 &  202 &  174
         &   995 &    700 &   1420 & 309 &  13.2 \\
 290&350 &   835 &  395 & 1500 &   4500 &  412 &  560
         &  1080 &    335 &   2660 & 510 &  11.6 \\
   \tableline
   0&350 & 16760 & 2465 & 5100 & 665000 & 1786 & 6020
         &  9710 &1085000 & 880000 & 1658 & 115.3 \\
   2&350 &  8180 & 1453 & 4520 &  29750 & 1716 & 1587
         &  8710 &   9830 &  19770 & 1580 & 107.7 \\
 \multicolumn{2}{c|}{$\chi^{2}/N_{\rm data}$} & 6.1 & 1.9 & 3.8
         & 20 & 2.0 & 1.9 & 6.4 & 7.1 & 13 & 1.9 & 1.0 \\
\end{tabular}
\label{chipot}
\end{table}

\mediumtext
\begin{table}
\caption{The difference ${\protect\scriptstyle\Delta}\chi^{2}$ (see
         text) in the 2--350 MeV energy range of the potential models
         using all potential phase shifts, or using one particular
         phase shift only.}
\begin{tabular}{r|rrrrrrrr}
        & \multicolumn{1}{c}{all}
        & \multicolumn{6}{c}{one particular phase shift} &         \\
  model & \multicolumn{1}{r}{phases} & ${^{1}S}_{0}$ & ${^{3}P}_{0}$
        & ${^{3}P}_{1}$ & ${^{3}P}_{2}-{^{3}F}_{2}$
        & ${^{1}D}_{2}$ & ${^{3}F}_{3}$ &  sum  \\
  \tableline
 Nijm78 & 1614 &  283  &  396  &    5  &  462  &
                  570  &  378  & 2094        \\
Paris80 & 1480 &  165  &  215  &  139  &  709  &
                  600  &  232  & 2060        \\
 Bonn89 & 1478 &  720  &  481  &   87  &  695  &
                  340  &   84  & 2407        \\
NijmRdl &  4.9 &  2.9  &  1.1  &  0.4  &  4.5  &
                --6.6  &--0.6  &  1.7        \\
\end{tabular}
\label{potwave}
\end{table}

\end{document}